\documentclass[prl,twocolumn,superscriptaddress,lengthcheck]{revtex4-2}%
\usepackage{amsmath,amssymb}
\usepackage{graphicx}
\usepackage[
    unicode,
    pdfauthor={Kotov et al.},
    colorlinks=true,
    urlcolor=blue,
    linkcolor=blue,
    citecolor=blue
]{hyperref}

\def\equationautorefname#1#2\null{Eq.#1(#2\null)}

\begin{document}
\title{Casimir-Lifshitz Theory for Cavity Modification of Ground-State Energy}

\newcommand{\Cham}{
    Department of Physics,
    Chalmers University of Technology,
    412~96, G\"oteborg, Sweden
}
\newcommand{\FTMC}{
    Departamento de Física Teórica de la Materia Condensada, Universidad Autónoma de Madrid, E-28049 Madrid, Spain
}
\newcommand{\IFIMAC}{
    Condensed Matter Physics Center (IFIMAC), Universidad Autónoma de Madrid, E-28049 Madrid, Spain
}

\author{Oleg V. Kotov}
\email{oleg.kotov@uam.es}
\affiliation{\Cham}
\affiliation{\FTMC}
\affiliation{\IFIMAC}

\author{Johannes Feist}
\affiliation{\FTMC}
\affiliation{\IFIMAC}

\author{Francisco J. García-Vidal}
\affiliation{\FTMC}
\affiliation{\IFIMAC}

\author{Timur O. Shegai}
\email{timurs@chalmers.se}
\affiliation{\Cham}

%\vspace{2cm}

\begin{abstract}  % ≤ 600 characters for PRL

A theory for ground-state modifications of matter embedded in a Fabry-Perot cavity and whose excitations are described as harmonic oscillators is presented. Based on Lifshitz's theory for vacuum energy and employing a Lorentz model for the material permittivity, a nonperturbative macroscopic QED model that accounts for the infinite number of cavity modes with a continuum of their wave vectors was built. Differences from the commonly used single-mode Hopfield Hamiltonian are revealed. The nonresonant role of polaritons in the ground-state energy shift is also demonstrated, showing that the cavity effect is mainly caused by static screening occurring at very low frequencies. The theory allows for a straightforward incorporation of losses and temperature effects. 

\end{abstract}
%\date{\today}

% Length for PRL letter Letter 3,750 words includ. figs
\maketitle

%Introduction

\textit{Introduction}. Vacuum-induced modifications of molecular properties in a ``dark'' cavity have recently attracted considerable attention~\cite{ebbesen2021}. It is claimed that strong coupling (SC) of electromagnetic (EM) modes to material excitations can modify a range of material properties, including chemical reaction rates~\cite{ebbesen2016,ebbesen2019,simpkins2023,GomezRivas2024,brawley2025vibrational}, dielectric constants~\cite{fukushima2022}, work functions~\cite{ebbesen2013}, phase transitions~\cite{ebbesen2014,jarc2023cavity}, ferromagnetism~\cite{ebbesen2021Ferro}, and (super) conductivity~\cite{ebbesen2015,sentef2018cavity,ebbesen2019Super,moddel2021,faist2022,kumar2024extraordinary,keren2025cavity}.
It has long been suggested that ultrastrong coupling (USC) of matter to individual cavity modes can modify the system's ground state~\cite{ciuti2005,fornRev2019,friskRev2019}.
Vibropolaritonic chemistry is typically realized in Fabry-Perot (FP) cavities by the effect of collective strong coupling with a very large number of molecules and without external driving, i.e., at thermodynamic equilibrium, suggesting that USC effects could be responsible. However, available theoretical approaches do not explain the observed changes under these conditions~\cite{Li2022Molecular, Fregoni2022Perspective, Xiang2024}. There is thus a clear need for a theoretical approach that links USC to polariton chemistry and can incorporate the experimentally relevant conditions.

A wide range of quantum optical Hamiltonian approaches has been employed to account for ground-state modifications in the SC and USC regimes~\cite{mandalRev2023,ruggRev2023,Foley2023}. However, these methods are typically restricted to a single~\cite{ciuti2005} or a few~\cite{mandal2023,herrera2024,Sidler2024} EM modes with a fixed wave vector, which severely limits their applicability. Furthermore, they treat a cavity as creating new EM modes, instead of taking into account that it primarily rearranges existing ones. The need to subtract the unmodified background leads us to a framework of Casimir-Lifshitz dispersion forces~\cite{lifshitz1956,DLP1961}, which is based on such subtraction and accounts for the infinite number of cavity modes with a continuum of their wave vectors. Starting from the full cavity Hopfield Hamiltonian we derive the exact ground-state energy of oscillators in a cavity matching it to the Casimir-Lifshitz energy with Lorentz permittivity. We regularize infinite sums of polaritonic zero-point energies (ZPEs) by using a Wick rotation to the imaginary frequency axis and subtracting the infinite free-space ZPE. Perturbatively, a similar approach has been previously applied for a single oscillator in a cavity~\cite{barton1970,rabl2023}, leading to the Casimir-Polder energy~\cite{casimirPolder1948,buhmann1,rabl2023}. However, summing up Casimir-Polder interactions for an ensemble of molecules is significantly more challenging. By contrast, the Lifshitz approach, which treats molecules as a homogeneous medium described by a Lorentz dielectric function, allows for an exact, \textit{nonperturbative}, and cost-efficient calculation providing an accurate analytical approximation for the ground-state energy shift. Furthermore, the theory naturally accommodates arbitrary mirror materials, material losses, and finite-temperature effects.
%To evaluate vacuum-related effects in a cavity system comprising a large number of molecules, Lifshitz approach allows for an exact, \textit{non-perturbative}, and cost-efficient calculation of the vacuum energy shift by modeling the macroscopic medium as an ensemble of harmonic oscillators. It naturally accommodates arbitrary mirror materials, material losses, and finite-temperature effects. %The ground state of the coupled light-matter system can thus be described as a \textit{polaritonic} vacuum, which captures both the modification of matter’s ground state due to the electromagnetic environment and the back-action of matter on vacuum field fluctuations.
%By integrating the \textit{continuum} of modes of the system over imaginary frequencies using the \textcolor{blue}{Lifshitz-Lorentz} formalism, we obtain both an exact numerical solution and an accurate analytical approximation for the ground-state energy shift.
%By integrating the \textit{continuum} of modes of the system over imaginary frequencies, we obtain both an exact numerical solution and an accurate analytical approximation for the ground-state energy shift.
%Furthermore, the theory naturally accommodates arbitrary mirror materials, material losses, and finite-temperature effects.
A Wick rotation converts all resonant polaritonic features into monotonic functions so that the main contribution to the ground-state energy originates from small imaginary frequencies, which contain information about polaritons, but in a nonresonant way. 

\textit{Lorentz permittivity and QED Hopfield Hamiltonian.} We first consider an infinite resonant medium homogeneously and isotropically filled with atoms or molecules having the same resonant frequencies $\omega_0$ and relaxation rates $\gamma$. In a linear approximation, they can be treated as harmonic oscillators, and neglecting the interaction between them, the dielectric function of the entire medium for light wavelengths much larger than the distance between the oscillators can be described by the classical Lorentz permittivity: 
\vspace{-5pt}
\begin{equation}
 \varepsilon(\omega,\mathrm{g}) = 1 + \frac{f\omega_p^2}{\omega_0^2 - \omega^2 - i\omega\gamma}= 1 + \frac{4\mathrm{g}^2}{\omega_0^2- \omega^2 - i\omega\gamma}, 
 \label{eps}
\end{equation} 
where $\omega_{p}$ and $f$ are the collective plasma frequency of the oscillators and the oscillator strength, respectively, and $\mathrm{g}$ is a measure of the light-matter coupling. 

From a quantum perspective, these Lorentz materials in the lossless limit $\gamma\to 0$ can be described using the so-called QED Hopfield Hamiltonian~\cite{hopfield}:
\begin{multline}
\hat{H} =\sum_{k,\lambda}\biggl[\hbar \omega_{k}\left(\hat{a}_{k,\lambda}^\dagger \hat{a}_{k,\lambda} + \frac{1}{2} \right) + \hbar \omega_0 \left(\hat{b}_{k,\lambda}^\dagger \hat{b}_{k,\lambda} + \frac{1}{2} \right)
\\
+ \hbar\mathrm{g_C}\left( \hat{a}_{k,\lambda}^\dagger + \hat{a}_{k,\lambda} \right) \left( \hat{b}_{k,\lambda}^\dagger + \hat{b}_{k,\lambda} \right) + \frac{\hbar\mathrm{g_C^2}}{\omega_0} \left( \hat{a}_{k,\lambda}^\dagger + \hat{a}_{k,\lambda} \right)^2\biggr],
\label{Hoplike}
\end{multline}
where the sum is taken over the conserved wave vector $k$ and two polarizations $\lambda=(\mathit{p,s})$, while $\hat{a}_{k,\lambda}^\dagger$ and $\hat{b}_{k,\lambda}^\dagger$ are the creation operators of the free-space photons ($\omega_{k}=c k$) and of the collective matter excitations with dispersionless frequency $\omega_0$, respectively. Importantly, the light-matter interaction strength in this minimal-coupling description is given by $\mathrm{g^2_C}=\mathrm{g}^2\, \omega_0/\omega_{k}$, and thus is $k$-dependent, such that the Lorentz permittivity is recovered when calculating the normal modes (\textit{bulk polaritons}) of the QED Hopfield Hamiltonian (see Supplemental Material (SM) \textcolor{black}{Sec.~I}~\cite{Suppl}).
 
\begin{figure}[tp]
\includegraphics[width=0.45\textwidth]{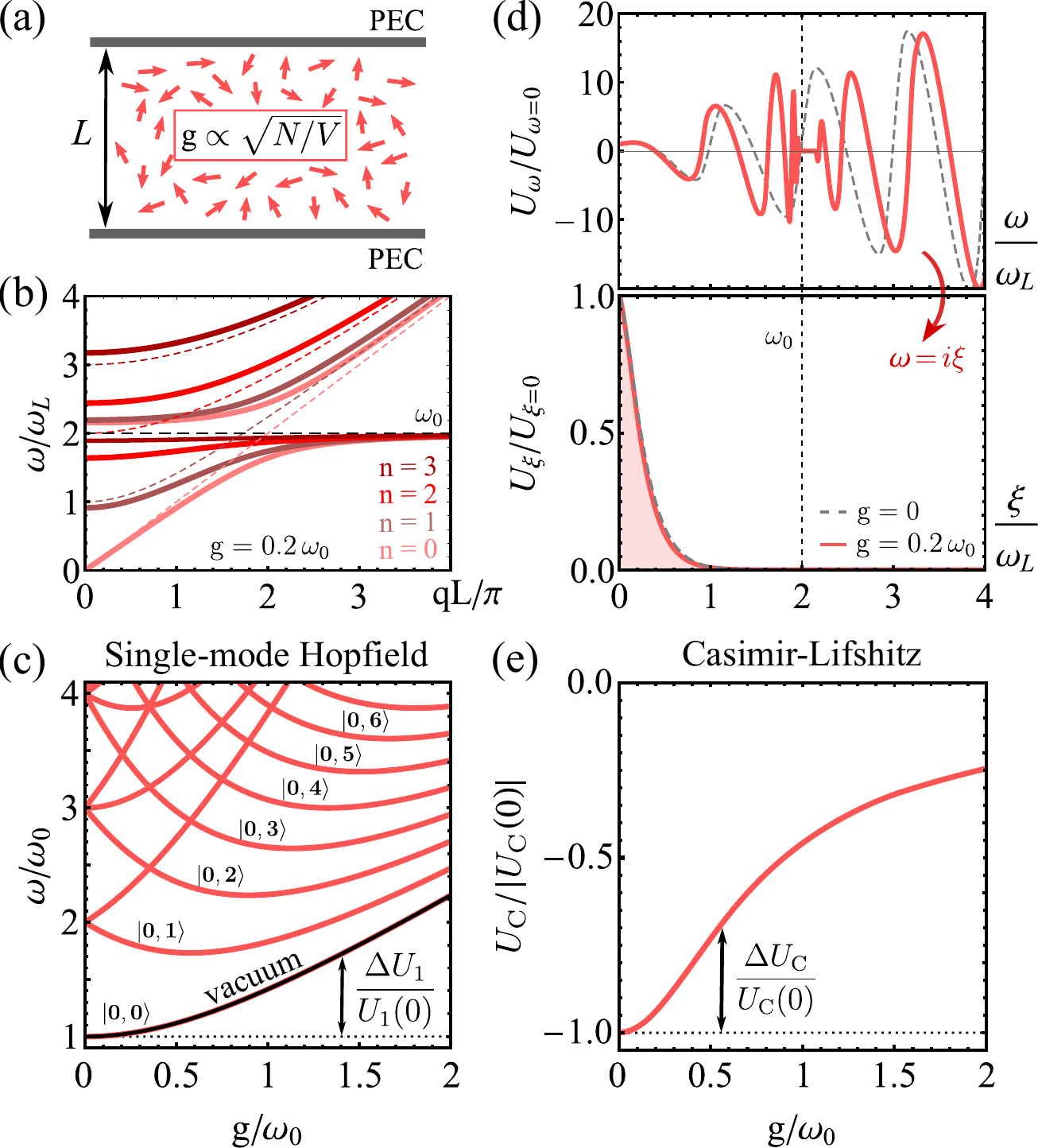}
\caption{\footnotesize{(a) Sketch of a resonant system consisting of identical harmonic oscillators strongly coupled to the vacuum EM modes in a PEC FP cavity. (b) Cavity polaritons dispersion at $\mathrm{g}=0.2\omega_0$ for $\omega_0=2\omega_L$; the empty cavity modes are shown in dashed lines. (c) Single-mode Hopfield Hamiltonian spectrum for the cavity polaritons at normal incidence, fixed polarization and zero detuning ($\omega_0=\omega_L$) as a function of $\mathrm{g}$. (d) The real (top) and imaginary (bottom) frequency dependence of the Lifshitz integrand at $T=0$ in the empty cavity (gray dashed) and polaritonic cavity (red solid). (e) Casimir-Lifshitz energy at $T=0$ and $\omega_0=\omega_L$, normalized to the empty cavity case as a function of the coupling energy, $\mathrm{g}$.}} 
\label{Fig1}
\end{figure}

\textit{Cavity polaritons.}  We are interested in analyzing the ground-state energy of a system consisting of the material described by a Lorentz permittivity embedded in an FP cavity. We first analyze the case of FP mirrors made of a perfect electrical conductor (PEC), see \autoref{Fig1}(a). For calculating the normal modes of the system, i.e., \textit{cavity polaritons}, we can write a Hopfield-like Hamiltonian~\cite{ciuti2005} similar to that of \autoref{Hoplike}, but instead of the unbounded light frequency $ck$, there appear discrete bands of cavity modes, $\omega_{q,n}=c \sqrt{q^2+(\pi n/L)^2}$, with $n$ denoting the out-of-plane mode number and $q$ being the (continuous) in-plane momentum. In \autoref{Fig1}(b), we render the dispersion of the cavity polaritons supported by an FP cavity for the case of $\mathrm{g}=0.2 \omega_0$ and $\omega_0=2\omega_L$, where $\omega_L\equiv \pi c/L$. At large enough $\mathrm{g}$, many cavity polariton branches arise, which account for the coupling of material excitations to multiple cavity modes.
%Sufficiently large $\Re\varepsilon(\mathrm{g})$ provides a whole series of lower polaritons, which are suppressed only near the resonant frequency $\omega_0$ by large $\Im\varepsilon(\mathrm{g})$.
 
As discussed below, a single-mode approximation to the cavity Hopfield Hamiltonian is often utilized when dealing with cavity polaritons. Moreover, in some cases, the dispersion of the fundamental ($n=1$) mode of the cavity is neglected ($q=0$). In this case, the energy levels of the single-mode Hopfield Hamiltonian are easily written as: $\omega_{l,m}=(l+\frac12) \omega_1^{+}+ (m+\frac12)\omega_1^{-}$, where $l$ and $m$ are non-negative integers, and  $\omega_1^{\pm}$ are the frequencies of the two polaritons formed by the coupling between the fundamental mode $\omega_L$ and $\omega_0$, compare \autoref{Fig1}(c). 
  
\textit{Ground-state energy change: Casimir-Lifshitz versus single-mode Hopfield Hamiltonian}.
The ground-state energy, i.e., ZPE of a material embedded in an FP cavity can be found from the full cavity Hopfield-like Hamiltonian and expressed as the half-sum of the FP cavity polaritons:
\begin{equation}
    U(\mathrm{g},\mathit{L}) = \frac{1}{2} \sum_{a,q,\lambda} \hbar\omega_{\lambda}(a,q,\mathrm{g}, L), \label{ZPE}
\end{equation}
where the sum involves the discrete polaritonic state index $a$ (with typically two polaritons for each mode index $n$), the in-plane wave vectors $q$, and the polarization $\lambda$. The sum over the conserved wave vectors $q$ can be expressed through an integral with the corresponding density of states. \autoref{ZPE} already accounts for both photonic and matter oscillators' ZPEs, as well as their coupling term. However, to obtain a finite value from this infinite energy, one needs to subtract the ZPE of the uncoupled system. 
Direct subtraction, $U(\mathrm{g}, \textit{L})-U(0, \textit{L})$, may still diverge (see SM \textcolor{black}{Sec.~II}~\cite{Suppl}), while Casimir-type subtraction of the ZPE within the same volume but using the continuous dispersion relation obtained outside a cavity, $U_\mathrm{C}(\mathrm{g},L) = U(\mathrm{g},L) - U_{\infty}(\mathrm{g},L)$, leads to a well-defined energy. Therefore, in order to get the cavity-induced ground-state energy shift due to light-matter coupling, we calculate the difference between Casimir energies as the coupling is turned on: $\Delta U_{\rm C}=U_{\rm C}(\mathrm{g},L)-U_{\rm C}(0,L)$.

A direct calculation of the ground-state energies by summing over the real frequencies of all cavity polaritons, as written in \autoref{ZPE}, does not converge. The key approach that we take to circumvent this problem was developed by Barash and Ginzburg~\cite{barash1972,barash1975,barash1989}, which also naturally allows for treating dissipative matter oscillators [$\gamma>0$ in \autoref{eps}] and nonperfect mirrors. This approach establishes a fundamental connection between the equilibrium average energy (ZPE at zero temperature) of a system of damped oscillators and the dispersion relation for the eigenmodes of that system. Barash-Ginzburg theory proves that the polaritonic ZPE in a cavity, $U_{\rm C}(\mathrm{g},L)$, corresponds to the standard expression for the Lifshitz energy per unit area $S$ (see SM~\cite{Suppl}), which at zero temperature is given by~\cite{barash1989,bordag}

\begin{equation}
    \frac{U_{\rm C}(\mathrm{g},L)}{S}=\frac{\hbar}{4\pi^2}\int\limits_{0}^{\infty} \! q\mathrm{d}q \int\limits_{0}^{\infty}\!\mathrm{d}\xi \sum_{\lambda=p,s}
    \ln(1-r_\lambda^-r_\lambda^+\mathit{e}^{-2k_z L}),
\label{Lif} 
\end{equation}
where $k_z =\sqrt{q^2+\varepsilon(i\xi,\mathrm{g})\xi^2/c^2}$ and $r_\lambda^\pm(q,i\xi,\mathrm{g})$ are the Fresnel reflection coefficients of the top and bottom mirrors (including the substrate). Notice that, for an empty PEC cavity ($r_\lambda^-r_\lambda^+=1$, $\varepsilon=1$),  \autoref{Lif} gives the well-known Casimir energy $U_{\rm C}(0,L)=-\hbar c \pi^2 L^{-3}/720$~\cite{casimir1948}. In \autoref{Lif}, the integration is performed over imaginary frequencies $\omega=i\xi$ \textcolor{black}{(see SM Sec.~III.A~\cite{Suppl})}, which, together with the subtraction of the cavity-free limit, eliminates the divergence of \autoref{ZPE}. Importantly, \autoref{Lif} contains the Lorentz permittivity, \autoref{eps}, since it is derived from \autoref{ZPE} with cavity polaritons whose dispersion relation is governed by this specific permittivity.
% Note that this Casimir-Lifshitz energy can be considered as that obtained from the full cavity Hopfield Hamiltonian after properly evaluating the two sums (cavity modes and wave vectors) in \autoref{ZPE}.
%\textcolor{blue}{The summation over all photon states leads to the asymmetric contribution of the material and photonic parts in \autoref{Lif}.}

At real frequencies, the Lifshitz integrand $U_\omega$ (given by \autoref{Lif} evaluated after wave vector integration but before frequency integration) behaves similarly to the local density of photonic states in the cavity. It exhibits periodic sign changes, has a polaritonic gap, and (for PEC mirrors) grows without limit with frequency~\cite{Rodriguez2013}, see top panel of \autoref{Fig1}(d). This behavior explains why a direct calculation of $U_{\rm C}(\mathrm{g},\textit{L})$ based on \autoref{ZPE} does not converge and why including more cavity modes within a few-mode approximation to the Hopfield Hamiltonian can change the answer even qualitatively without necessarily improving agreement with the correct result~\cite{baranov2020,mandal2023}. 

Wick rotation to imaginary frequencies eliminates not only the divergence but also all resonant features, making the integrand $U_{\xi}$ smooth, monotonic, and rapidly decaying, compare bottom panel of \autoref{Fig1}(d). Although it encompasses all of the information about the polaritons, their visual impact compared to $\rm g=0$ is barely noticeable. The integrand decays rapidly from its maximum at $\xi=0$, where for PEC at $T=0$ it is given by $U_{\xi=0} = -\zeta(3)\big/(8 L^2\pi^2)$, with approximately 99\% of the total energy originating from imaginary frequencies smaller than the fundamental cavity mode frequency $\omega_L$.
%Moreover, as $\mathrm{g}$ increases, the integrand decays even more rapidly, thereby increasing the relative contribution of the $i0$ frequency.
The nonresonant behavior of $U_{\xi}$ leads to a monotonic dependence of the Casimir-Lifshitz energy $U_{\rm C}$ on $\mathrm{g}$, as shown in \autoref{Fig1}(e).

For PEC mirrors at $T=0$, it is feasible to obtain an analytical approximation for the cavity-induced change of the ground-state energy. By taking the screening factor in the static limit, $1\Big/\sqrt{\varepsilon(i\xi=0,\mathrm{g})}$, out of the integral in \autoref{Lif}, the relative Casimir energy change within what we call \textit{static screening approximation} (SSA) can be written as (see SM \textcolor{black}{Sec.~IV}~\cite{Suppl}):
\begin{equation}
\frac{\Delta U_{\rm C}}{U_{\rm C}(0)}\Bigg|_{\rm stat.}\approx1-\frac{1}{\sqrt{\varepsilon\left(0,\mathrm{g}\right)}}=1-\frac{1}{\sqrt{1+4\mathrm{g}^2/\omega_0^2}}. \label{SSA}
\end{equation}

\begin{figure}[tp]
\includegraphics[width=0.47\textwidth]{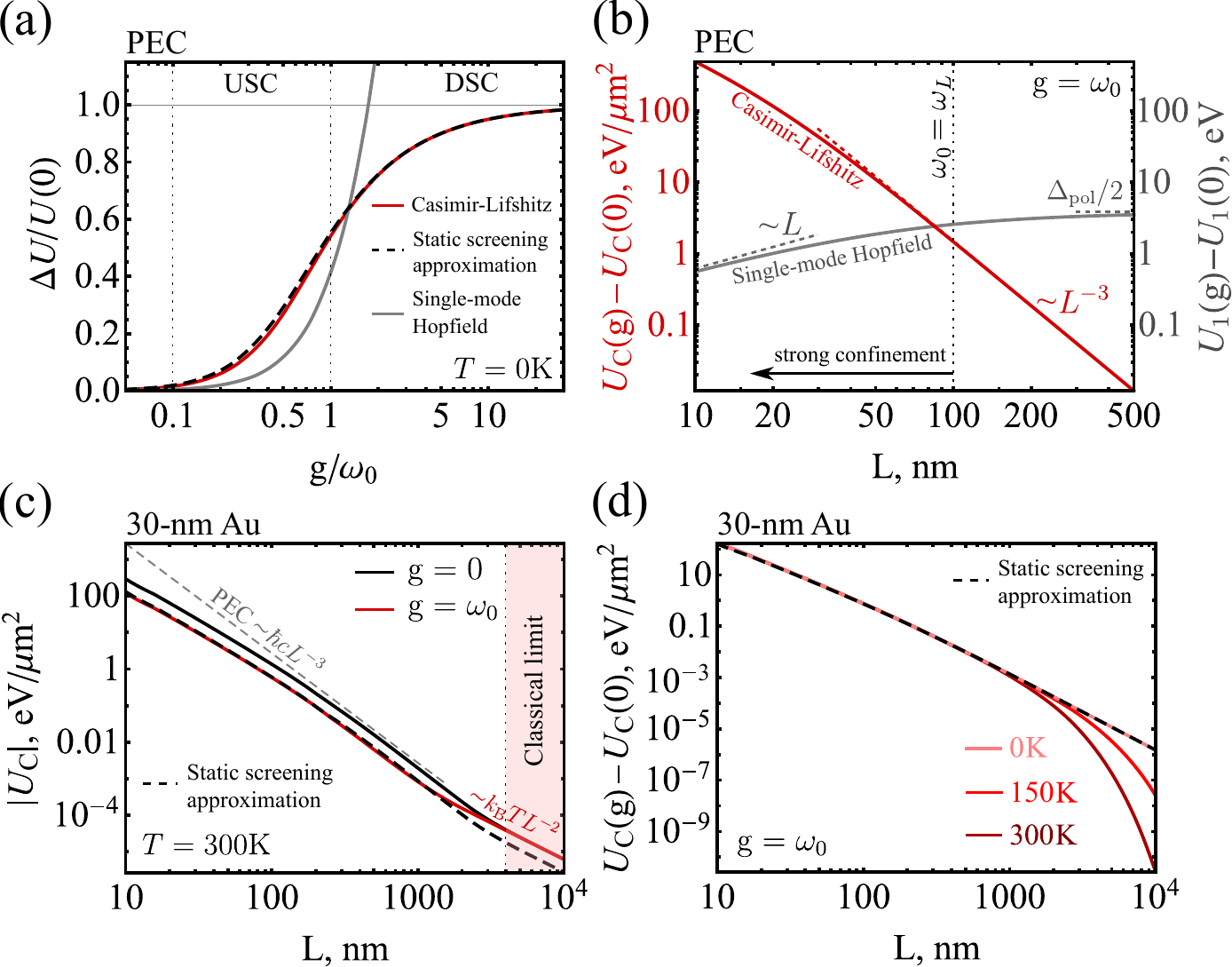}
\caption{\footnotesize{(a) Relative change of the ground-state energy vs. coupling $\mathrm{g}$ in a PEC cavity at $T=0$K and $L=100$ nm according to Lifshitz (red) and single-mode Hopfield (gray) approaches. The static screening approximation (SSA) is shown in dashed lines. USC and deep strong coupling (DSC) denote ultrastrong and deep stong coupling regimes, respectively. (b) Absolute change of the ground-state energy vs $L$ at $T=0$K, with the left axis for Lifshitz and the right one for single-mode Hopfield solutions. (c) Casimir-Lifshitz energy of the cavity with 30-nm gold mirrors on a glass substrate at $T=300$K without (black) and with (red) a medium (with bulk coupling $\mathrm{g} =\omega_0$). SSA results are depicted by dashed lines. (d) Absolute change in ground-state energy at different temperatures. SSA works perfectly at $T=0$K, as expected. In all the plots $\omega_0$ is tuned to the main mode of $L=100$ nm cavity ($\omega_0=\omega_{L=100{\space \rm nm}}$).
}}
\label{Fig2}
\end{figure}

As shown in \autoref{Fig2}(a), at $T=0$ the SSA reproduces the exact Lifshitz solution extremely well for the whole range of $\mathrm{g}/\omega_0$, confirming the key role of the zero-frequency limit for the ground-state energy shift.
This is reminiscent of earlier results showing that the cavity-mediated interaction between low-energy excitations reduces to the electrostatic limit~\cite{pantazopoulos2024electrostatic,SanchezMartinez2024General,polini2024,polini2025}.

As commented above, many works have employed a single-mode Hopfield Hamiltonian to study cavity-induced changes in ground-state energy. Within this single-mode approximation, the energy shift can be calculated as the difference between the polaritonic ZPE and the ZPE of the uncoupled system~\cite{ciuti2005,baranov2020}: $\Delta U_1=\hbar\left(\omega_1^{+} +\omega_1^{-} -\omega_0-\omega_L\right)/2$. Unlike the Casimir-Lifshitz energy, the single-mode Hopfield ZPE does not contain the free-space subtraction and does not scale with the mirror area as no integration over parallel wave vectors is performed. 
%\textcolor{black}{Such single-mode models thus treat a cavity as creating new EM modes, instead of taking into account that it primarily rearranges existing ones.}
However, when considering the relative change of the ground state energy, a quantitative comparison between Casimir-Lifshitz and single-mode Hopfield energy shifts becomes possible, as the area dependence in the Casimir energy cancels out. In the limit $\mathrm{g}\ll\omega_0$, the single-mode Hopfield relative energy changes simplifies to
\begin{equation}
\frac{\Delta U_1}{U_1(0)}=\frac{2{\mathrm{g}}^2}{(\omega_0+\omega_L)^2}\,+\,O(\rm g^4). \label{UHop}
\end{equation}

For a PEC cavity at $T=0$, a  detailed comparison between the results of the Lifshitz-Lorentz approach and those obtained with the single-mode Hopfield Hamiltonian is rendered in panels (a) and (b) of \autoref{Fig2}. For the relative change of the ground-state energy as a function of $\mathrm{g}$, at $\mathrm{g}\ll\omega_0$, both curves grow quadratically but with very different prefactors. 
%Only in the limit of very large cavities ($L \rightarrow \infty, \omega_L \rightarrow 0$), both approaches scale as $2\mathrm{g}^2/\omega_0^2$.  
However, already in the USC regime (at the inflection point $\mathrm{g}\approx0.4\omega_0$), the Lifshitz curve begins to bend toward saturation at $\mathrm{g}\gg\omega_0$, whereas the single-mode solution in the DSC regime shows unbounded linear growth [\autoref{Fig2}(a)]. The Lifshitz-Lorentz approach provides a physically meaningful saturation at the high-$\mathrm{g}$ limit, where $\varepsilon(\mathrm{g})$ becomes so large that it fully screens the Lifshitz energy, effectively suppressing any further modifications. This limit corresponds to a very large number of oscillators, analogous to the thermodynamic limit $N\gg1$ in microscopic analyses~\cite{iorsh2023}. In this regime, the ground state of the harmonic oscillators cannot be modified through coupling to vacuum modes by more than the initial vacuum energy contained within them, $\Delta U/U(0)\leq1$. 

For polaritonic chemistry phenomena, it is also relevant to know the absolute change in ground-state energy. Comparing $\Delta U$ from both approaches on the same plot is valuable, as it reveals their qualitatively different behavior with respect to $L$. Unlike the relative energy difference, $\Delta U$ obeys a fundamentally different $L$ scaling. The single-mode Hopfield ZPE increases linearly with $L$ in tightly confined cavities ($L\ll\pi c/\omega_0$) and saturates at a constant value determined by the polaritonic gap $\Delta_{\rm pol}/2$ in large ones [\autoref{Fig2}(b)],
\textcolor{black}{which is unphysical, since $L \rightarrow \infty$ corresponds to the absence of a cavity (see SM Sec.~III.C~\cite{Suppl})}. By contrast, the Lifshitz energy rapidly decreases with increasing $L$, reaching its static limit scaling with $L^{-3}$ already around $L=50$~nm. 

\textit{Temperature effects and non-PEC mirrors}. Lifshitz's formalism, in contrast to the single-mode Hopfield Hamiltonian, can easily incorporate the effects of temperature and nonperfect cavity mirrors made of real materials. At a finite temperature $T$, the integral over imaginary frequencies in \autoref{Lif} is replaced by a sum over Matsubara frequencies $\xi_j=2\pi jk_{\rm B}T/\hbar$, where $k_{\rm B}$ is the Boltzmann constant, $j=0,1,2,\ldots$, and the term with $j = 0$ is multiplied by $1/2$. Non-PEC mirrors can be accounted for in \autoref{Lif} by evaluating their associated Fresnel coefficients, $r^{\pm}_{p,s}$. At sufficiently high temperatures or distances $L$, classical thermal fluctuations completely dominate quantum ones \textcolor{black}{(see SM Sec.~III.B~\cite{Suppl})}. Instead of the Casimir power law $\propto\hbar cL^{-3}$, in the classical limit the energy scales as $\propto k_{\rm B}TL^{-2}$~\cite{lifshitz1956,mehra1967,schwinger1978}. In this limit, only the $\xi=0$ contribution remains. Moreover, in this effectively electrostatic limit, the reflection becomes perfect not only for PECs but even for realistic Drude mirrors (although the reflection for s polarization vanishes for $\xi=0$), and the contribution of $\varepsilon(\mathit{i\xi},\rm g)$ completely vanishes. This can be clearly seen in Figs.~\ref{Fig2}(c),(d), which show the Lifshitz energy for gold mirrors (Drude model) at different temperatures. $U_{\rm C}(0)$ deviates from the PEC $L^{-3}$ scaling at $L<1\mu$m but reaches a similar classical limit $L^{-2}$ scaling at $L\approx4\mu$m. On the other hand, $U_{\rm C}(\mathrm{g})$ merges with $U_{\rm C}(0)$ when they both reach the classical limit [\autoref{Fig2}(c)]. This shows a complete absence of vacuum-induced energy shifts in mid-infrared FP cavities at room temperature [\autoref{Fig2}(d)]. %\textcolor{blue}{Note that for gold mirrors, SSA perfectly reproduce the $L$-behavior of $U_{\rm C}(g)$ until the temperature effects start to appear.} 

\begin{figure}[tp]
\includegraphics[width=0.3\textwidth]{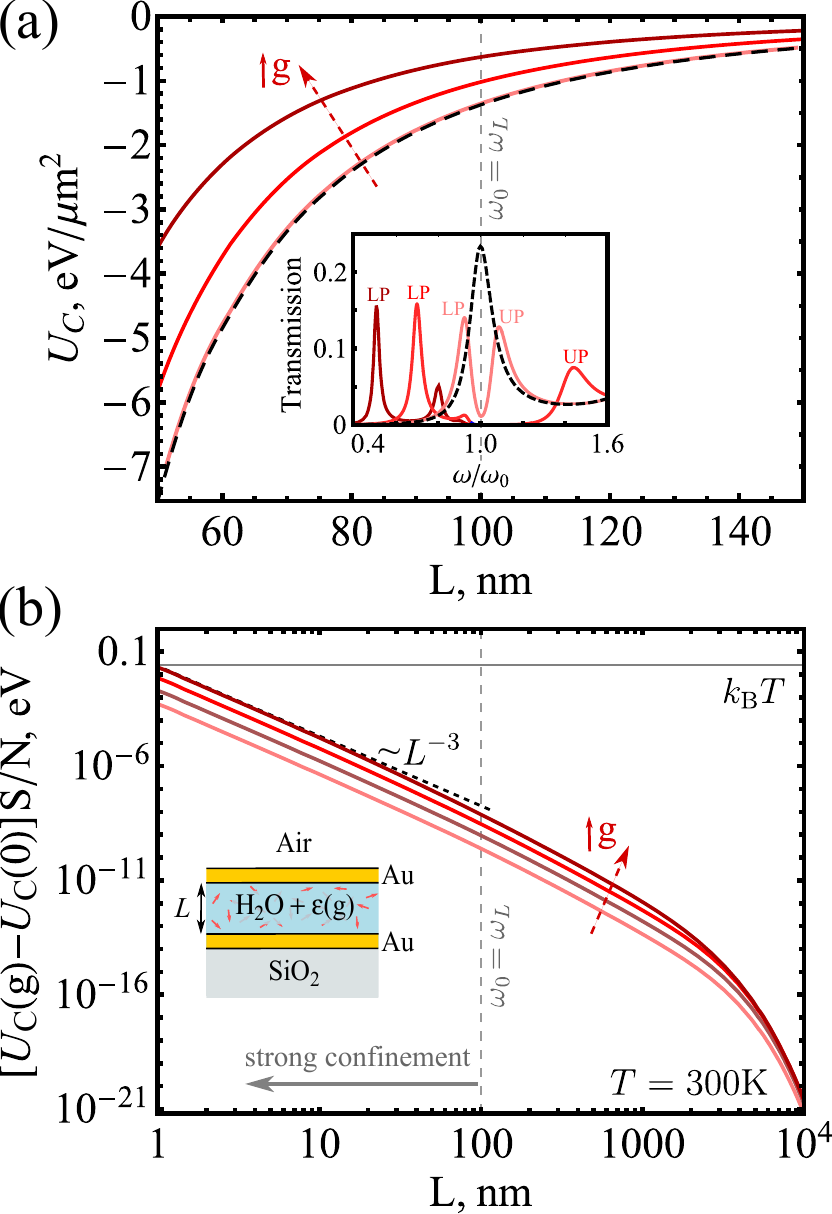}
\caption{\footnotesize{(a) Non-resonant behavior of the Casimir-Lifshitz energy for a gold cavity on a glass substrate filled with molecules floating in water at room temperature (see the sketch). Different concentrations of molecules leads to $\mathrm{g}/\omega_0$ varying from 0 (black dashed) to 0.1, 0.5, 1 (light to dark red), where $\omega_0$ equals to the main mode of 100-nm-cavity. The inset shows the corresponding transmission spectrum with the resonant splittings having USC and DSC features. (b) For the same system, the change of the Lifshitz energy per molecule, showing a similar scaling law as non-retarded Casimir-Polder energy at small $L$ and vanishing of the energy shift at large $L$}.
}
\label{Fig3}
\end{figure}

\textit{Casimir-Lifshitz energy and cavity polaritons}. \autoref{Fig3}(a) shows the absolute value of Casimir-Lifshitz energy for varying couplings, calculated with \autoref{Lif} for realistic FP cavities with gold mirrors and molecular oscillators in water. When the cavity is tuned to the oscillator resonance, $\omega_0=\omega_L$, we do not observe a resonant behavior of the vacuum energy, similar to the nonresonant effect of Casimir-Polder shifts on chemical reactions~\cite{galego2019}. Instead, it is gradually suppressed with increasing $\mathrm{g}$, whereas the transmission spectra display typical polariton splitting with USC and even DSC features appearing at progressively increasing couplings, see inset in \autoref{Fig3}(a). Thus, polaritons are indeed present, but they do not exert a resonant effect on the ground-state energy. \textcolor{black}{This prediction could be tested experimentally, e.g., by extracting the Casimir energy from the total equilibrium potential of self-assembled microcavities~\cite{munkhbat2021,kuccukoz2024,hovskova2025} or by examining the corresponding pressure using a liquid-phase AFM~\cite{biggs1994,munday2007,munday2009} (see SM Sec.~V~\cite{Suppl}). As a simple illustrative estimate, we use \autoref{SSA}, which at $\mathrm{g}\ll\omega_0$ yields $\Delta U/U(0)\approx2\mathrm{g}^2/\omega_0^2$, thus leading to a $2\%$ relative change at $\mathrm{g}=0.1\omega_0$.}

Given that chemical processes usually occur at the level of individual molecules, a key characteristic of polaritonic chemistry is the ground-state change per molecule~\cite{joel2018}. Assuming that all $N$ molecules contribute independently, we can simply divide the Casimir-Lifshitz energy by the number of molecules. For gold mirrors, the Lifshitz energy per molecule in the van der Waals limit is $U_{\rm C} S/N = U_{\rm C}/(\rho L)\propto L^{-3}$, where $\rho$ is the concentration of the molecules. The same scaling law is known for the nonretarded Casimir-Polder (London) energy~\cite{buhmann2}. Notice that even in extremely small nanocavities ($L<10$ nm) the energy corrections in the USC and even DSC regimes are smaller than $k_{\rm B}T$ at room temperature, see \autoref{Fig3}(b). 

To conclude, using Barash-Ginzburg theory to regularize the infinite sum of polaritonic zero-point energies, we have derived the exact ground-state energy change of a material characterized by a Lorentz permittivity when embedded in a Fabry-Perot cavity. In this way, our Casimir-Lifshitz calculation provides the ground-state energy associated with the full cavity QED Hopfield Hamiltonian.
%\textcolor{blue}{This energy reflects all Casimir features: $L^{-3}$ scaling, non-resonant detuning behavior, medium screening with its complete absence at high temperatures in large cavities.}
We show that the cavity-modification is mainly governed by the quasistatic response and that cavity polaritons do not exert a resonant effect on the ground-state energy. We have compared the results of this full calculation with those obtained with the commonly used single-mode Hopfield Hamiltonian, showing the severe limitations of this approach. We have also analyzed temperature effects and incorporated mirror losses. To test our findings, we suggest using Casimir measurements in Fabry-Perot cavities to experimentally probe ground-state modifications, thus bridging Casimir and polariton physics.

\acknowledgments{The authors acknowledge fruitful discussions with \mbox{D. G. Baranov}, \mbox{F. Lindel}, \mbox{S. Y. Buhmann}, \mbox{R. Podgornik}, and \mbox{V. M. Mostepanenko}. This work has been supported by the Swedish Research Council (VR project, Grant No. 202203347), the Knut and Alice Wallenberg Foundation (Grant No. 2019.0140), Olle Engkvist Foundation (Grant No. 211-0063), and Chalmers Area of Advance Nano. We also acknowledge financial support by the Spanish Ministerio de Ciencia y Universidades---Agencia Estatal de Investigación through grants PID2021-125894NB-I00, PID2024-161142NB-I00, EUR2023-143478, and CEX2023-001316-M through the María de Maeztu program for Units of Excellence in R\&D.}

\bibliography{arxivbib}

\onecolumngrid
\clearpage
\begin{center}
{
{\large \textbf{Supplemental Material: Casimir-Lifshitz theory for cavity-modification of ground-state energy}}
}
\thispagestyle{empty}

\vspace{0.15in}
Oleg V. Kotov\textsuperscript{1,2,3,\hyperlink{m1}{$^\ast$}}, Johannes Feist\textsuperscript{2,3}, Francisco J. García-Vidal\textsuperscript{2,3}, and Timur O. Shegai\textsuperscript{1,\hyperlink{m2}{$\dagger$}}

\vspace{0.075in}
\small\textit{\textsuperscript{1}Department of Physics,
    Chalmers University of Technology,
    412~96, G\"oteborg, Sweden}

\small\textit{\textsuperscript{2}Departamento de Física Teórica de la Materia Condensada, Universidad Autónoma de Madrid, E-28049 Madrid, Spain}

\small\textit{\textsuperscript{3}Condensed Matter Physics Center (IFIMAC), Universidad Autónoma de Madrid, E-28049 Madrid, Spain}

\hypertarget{m1}{$^\ast$} 
\href{mailto:oleg.kotov@uam.es}{\small oleg.kotov@uam.es} , \hypertarget{m2}{$^\dagger$}\href{mailto:timurs@chalmers.se}{\small timurs@chalmers.se} 

\end{center}

\section{I. Lorentz permittivity and Hopfield polaritons}
Due to the translational invariance in an unbounded medium and the independence of polarizations, it is sufficient to work with a single wavevector and polarization. Then the normal modes (bulk polaritons) of the Hopfield Hamiltonian, Eq.~(2) in the main text, can be found from the eigenvalue equation:
\begin{equation}
\left(\omega^2-\omega_k^2\right)\left(\omega^2-\omega_0^2\right)=4\mathrm{g_C}^2\omega^2\omega_k\big/\omega_{\rm 0},\label{S1}
\end{equation}
where the bare coupling strength with the vacuum field in the Coulomb gauge $\mathrm{g_C}$ is expressed through the polarizability $\alpha$ as follows~\cite{hopfield}: $\mathrm{g_C}=\omega_0^{3/2}\sqrt{\pi\alpha/\omega_k}$. On the other hand, the classical approach~\cite{canales2021} connects $\alpha$ with the oscillator strength $f$ of individual oscillators and their collective plasma frequency $\omega_{\rm p}$ in a way:
$4\pi\alpha\omega_0^2=f\omega_{\rm p}^2$, compare Eq.~(5) in Ref.~\cite{hopfield} and Eq.~(3) in Ref.~\cite{canales2021}. Thus, excluding $\alpha$, we obtain $f\omega_{\rm p}^2=4\mathrm{g_C}^2\omega_k\big/\omega_{\rm 0}\equiv4\rm g^2$. So, the eigenvalue equation (\ref{S1}) takes the form: $\left(\omega^2-\omega_k^2\right)\left(\omega^2-\omega_0^2\right)=4\mathrm{g}^2\omega^2$. Recalling the classical dispersion relation for bulk polaritons $\omega_k=ck=\omega\sqrt{\varepsilon(\omega,\rm g)}$, we obtain Eq.~(1) from the main text, but at $\gamma=0$. The damping term in Eq.~(1) can be proved by the Huttner-Barnett theory~\cite{huttner-barnett}, which accounts for the absorption and satisfies the Kramers-Kronig relations. 

The single-resonance Lorentz model under consideration can be straightforwardly extended to multiple resonances ($\varepsilon_\infty\neq1$) within both classical and quantum frameworks. However, in the absence of physical coupling adjustments, such extensions do not introduce qualitative changes, contributing only a $\rm g$-independent background. The Lorentz permittivity can be applied to a medium inside a cavity, assuming that the vast majority of oscillators are located much closer to their neighbors than to the mirrors, such that mirror-induced perturbations can be neglected. 

For a PEC cavity, one can write a Hopfield-like Hamiltonian~\cite{ciuti2005} similar to that of Eq.~(2) in the main text, but instead of the unbounded light frequency $ck$, a set of discrete cavity modes appears, $\omega_{q,n}=c \sqrt{q^2+(\pi n/L)^2}$, with $n$ denoting the mode number. In the single-mode approximation ($n=1$, $q=0$), the normal modes of this Hamiltonian are just two polaritons formed by the coupling between the fundamental mode $\omega_L=\pi c/L$ and $\omega_0$:
\begin{equation}
\omega_1^\pm(L,\mathrm{g})=\sqrt{\frac{\omega_L^2+\omega_0^2+4\mathrm{g}^2}{2}\pm\sqrt{\left(\frac{\omega_L^2+\omega_0^2+4\mathrm{g}^2}{2}\right)^2-\omega_0^2\omega_L^2}}.  \label{wpm}
\end{equation}
In this case, the difference between the
polaritonic ZPE and the ZPE of the uncoupled system is simply $\Delta U_1(\rm g)=\hbar \left(\omega_1^{+}(\rm g) +\omega_1^{-}(\rm g) -\omega_0-\omega_\textit{L}\right)/2$, which has the following limits as a function of $L$:
\begin{equation}
        \Delta U_1(\rm g)\rightarrow
        \begin{cases}
            {\rm g}^2L\big/\pi c,\, L\ll\pi c/\omega_0,
            \\
            \sqrt{\omega_0^2+4\rm{g}^2}\Big/2-\omega_0/2 = \Delta_{\rm pol}/2,\, L\gg\pi c/\omega_0,
\end{cases}
\end{equation}
where $\Delta_{\rm pol}$ is the polaritonic gap. The relative energy change, in the limit $\mathrm{g}\ll\omega_0$, simplifies to Eq.~(6) of the main text. In the static cavity mode limit $\omega_L\rightarrow0$ and for arbitrary $\rm g$, $\Delta U_1(\rm g)/{U_1(0)}=\sqrt{1+4\rm{g}^2/\omega_0^2}-1 = \Delta_{pol}/\omega_0$ looks similar to the expression obtained from the Lifshitz-Lorentz approach in the static screening approximation, compare with Eq.~(5) from the main text.

\section{II. Equivalence of the Hopfield model ZPE to the Lifshitz energy in a cavity}

Although the original Lifshitz formula was derived from the EM stress tensor specifically for the force~\cite{lifshitz1956}, the methods developed later to derive the corresponding energy allows one to prove the equivalence of the Casimir-Lifshitz energy in the FP cavity and Hopfield’s ZPE associated with the polaritonic modes in this cavity, i.e., correspondence between Eq.~(3) and Eq.~(4) in the main text. In a more general form, at non-zero temperature, Eq.~(3) defines the energy of the equilibrium fluctuating electromagnetic field (both thermal and quantum) with the corresponding free energy:
\begin{equation}
    U(\mathrm{g}, L) = \sum_{a,q} \frac{\hbar\omega_{a}(q,\mathrm{g}, L)}{2} \coth\frac{\hbar\omega_{a}(q,\mathrm{g}, L)}{2k_{\mathrm{B}}T}, \,\,\,\,\,\mathcal{F}(\mathrm{g}, L) = \sum_{a,q} k_{\mathrm{B}} T \ln\left[2 \sinh\frac{\hbar\omega_{a}(q,\mathrm{g}, L)}{2k_{\mathrm{B}}T}\right], \label{ZPE1}
\end{equation}
where the sum over $a$ involves all cavity polaritonic modes with both polarizations and arbitrary parallel wave vector $q$, the sum over which can be expressed through an integral with the corresponding density of states $\rho(q)$. The normal modes $\omega_{p,s}$ of the Fabry-Perot cavity are given by the dispersion relations corresponding to each polarization, $D_{p,s}=1-r_{p,s}^-(q,\omega,\mathrm{g}) r_{p,s}^+(q,\omega,\mathrm{g}) e^{-2L \sqrt{q^2 - \varepsilon(\omega,\mathrm{g}) \omega^2 / c^2}} = 0$, where $r_{p,s}^\pm(q,\omega,\mathrm{g})$ are the Fresnel reflection coefficients of the top and bottom mirrors. For thick metal as the first layer to the gap, typical expressions for semi-infinite mirrors can be used:
\begin{equation}
D_{p}(q,\omega,\mathrm{g}, L)=1-\frac{\left(\varepsilon_3 k_1-\varepsilon_1 k_3\right)\left(\varepsilon_3 k_2-\varepsilon_2 k_3\right)}{\left(\varepsilon_3 k_1+\varepsilon_1 k_3\right)\left(\varepsilon_3 k_2+\varepsilon_2 k_3\right)}e^{-2 k_3 L},\,\,\,\,\,\,\,\, D_{s}(q,\omega,\mathrm{g}, L)=1-\frac{\left( k_1- k_3\right)\left( k_2- k_3\right)}{\left( k_1+ k_3\right)\left( k_2+ k_3\right)}e^{-2 k_3 L},  \label{dispers}
\end{equation}
where $k_{1,2}=\sqrt{q^2-\varepsilon_{1,2}(\omega)\,\omega^2/c^2}$ are the transverse wave vectors inside the mirrors and $k_{3}=\sqrt{q^2-\varepsilon_{3}(\omega)\,\omega^2/c^2}$ with $\varepsilon_{3}(\omega)\equiv\varepsilon(\omega)$ in the medium between them.

The most common approach to derive the Lifshitz energy from the ZPE of cavity modes is based on the argument theorem~\cite{vankampen1968, parsegian1969}. However, this method regards the right-hand side of \autoref{ZPE1} as purely real, thus ignoring the absorption inherent in fluctuation processes due to the fluctuation-dissipation theorem. The challenge of incorporating absorption is that the eigenmodes $\omega_{p,s}$ in an absorbing and hence dispersive medium no longer represent an orthogonal basis; moreover, they become complex, and \autoref{ZPE1} no longer has a clear meaning of energy. To solve this problem, Barash and Ginzburg calculated the energy of the fluctuating field as the sum of the equilibrium energies of \textit{damped} harmonic oscillators driven by a fluctuation force $F(t)$, using the method of expansion in modes of an auxiliary system~\cite{barash1972,barash1975,barash1989}.
The expectation value of the equilibrium energy of the oscillators obeying equation $\ddot{x}+\alpha \dot{x}+\omega_0^2 x=F(t)/m$ is given by~\cite{milonni}:
\begin{equation}
U=\frac{\alpha}{2 \pi} \int\limits_{-\infty}^{\infty} \frac{\left(\omega^2+\omega_0^2\right) f(\omega, T)}{\left(\omega^2-\omega_0^2\right)^2+\alpha^2 \omega^2} d\omega,  \label{Eav}
\end{equation}
where $f\left(\omega, T\right)=\frac{\hbar\omega}{2} + \frac{\hbar\omega}{e^{\hbar\omega /  k_{\mathrm{B} T}} - 1} = \frac{\hbar\omega}{2} \coth\frac{\hbar\omega}{2k_{\mathrm{B}} T}$ is the Planck formula for the equilibrium energy of a single \textit{undamped} oscillator. However, it is impossible to sum the energy over all oscillator modes, since, as mentioned above, they do not form an orthogonal basis in the absorbing medium. This can be circumvented by introducing a formally nondispersive auxiliary system in which the permittivity depends on the frequency as a parameter, in accordance with the frequency dispersion of the system under study. For the auxiliary oscillator equation $\ddot{x}+\alpha\dot{x}\,\omega/\omega_1+\omega_0^2 x=0$ with $\omega_1$ being the parameter function of $\omega$, the solutions have the usual exponential form with $\omega_1^2=\omega_0^2-i\alpha\omega$. In terms of $\omega_1$ it is possible to rewrite \autoref{Eav} in the form~\cite{milonni,barash1972}:
\begin{align}
U &= -\frac{i}{\pi} \int\limits_{-\infty}^{\infty}d\omega \frac{\omega f(\omega, T)}{\omega_1^2(\omega)-\omega^2} + \frac{i}{2 \pi} \int\limits_{-\infty}^{\infty} d\omega \frac{-i \alpha f(\omega, T)}{\omega_1^2(\omega)-\omega^2} = -\frac{i}{\pi} \int\limits_{-\infty}^{\infty} d\omega \frac{\omega f(\omega, T)}{\omega_1^2(\omega)-\omega^2} + \frac{i}{2 \pi} \int\limits_{-\infty}^{\infty} d\omega \frac{f(\omega, T)\, \partial \omega_1^2(\omega) / \partial \omega}{\omega_1^2(\omega)-\omega^2} \nonumber \\
&= \frac{i}{2 \pi} \int\limits_{-\infty}^{\infty} d\omega f(\omega, T) \frac{\partial}{\partial \omega} \ln\left[\omega_1^2(\omega)-\omega^2\right].
\end{align}
Thanks to the orthogonality of the additional modes $\omega_a$~\cite{barash1975,barash1989}, we can sum over all additional oscillators:
\begin{align}
U &= \frac{i}{2 \pi} \sum_a \int\limits_{-\infty}^{\infty} d \omega f(\omega, T) \frac{\partial}{\partial \omega} \ln \left[\omega_a^2(\omega)-\omega^2\right] = \frac{i}{2 \pi} \int\limits_{-\infty}^{\infty} d \omega f(\omega, T) \frac{\partial}{\partial \omega} \ln \prod_a\left[\omega_a^2(\omega)-\omega^2\right] \nonumber
\\
&\rightarrow \frac{i}{2 \pi} \int\limits_{-\infty}^{\infty} d \omega f(\omega, T)\int d q\rho(q)\frac{\partial}{\partial \omega} \ln D(q, \omega), \label{Eav1}
\end{align}
where $\rho(q)$ is a density of states and $D(q, \omega)=0$ gives the dispersion relation for the modes $\omega_a$. The free energy corresponding to  \autoref{Eav1} is given by:
\begin{equation}
\mathcal{F} = -T \int \frac{U}{T^2} d T = \frac{\hbar}{2 \pi i}\mathcal{P}\int\limits_{-\infty}^{\infty} d \omega \frac{1}{e^{\hbar \omega /  k_{\rm B} T}-1} \int d q  \rho(q) \ln D(q, \omega),
\end{equation}
where the frequency integral at $\omega=0$ is taken in the sense of a principal value. In an equilibrium system, the zeros of $D(q, \omega)$, corresponding to the damped modes, should be in the lower part of the complex frequency plane and the poles of $f(\omega, T)$, the Matsubara frequencies $\xi_j=2\pi jk_{\rm B}T/\hbar$, are in the upper part. Closing the integration contour  in the upper half-plane, we finally obtain the generalized Planck formula for the case of dissipative equilibrium media:
\begin{equation}
 \mathcal{F}=k_{\rm B} T \sum_{j=0}^{\infty}{}^{\prime}\!\!\int dq\rho(q)  \ln D\left(q, i \xi_j\right) \label{Free}
\end{equation}
where the prime at the summation means that the term with $j=0$ is taken with a weight of $\frac12$. The dispersion function $D\left(q, i \xi_j\right)$ is determined within an arbitrary factor that does not depend on the inhomogeneity parameters. It must be normalized according to the physical meaning in a specific problem. For the vacuum energy in a Fabry-Perot cavity, the free energy must vanish at $ \mathcal{F}(L=\infty)\rightarrow0$. Accordingly, the relevant dispersions should be taken in the form of \autoref{dispers}, with the condition $D(L=\infty)\rightarrow1$. This corresponds to applying the normalization $D\left(q, i \xi_j,L\right)/D\left(q, i \xi_j,L=\infty\right)$, effectively subtracting the cavity-free contribution. Crucially, in the absence of such normalization (i.e., subtraction of the cavity-free limit), an undetermined additive constant remains in the energy, which generally results in a divergence. Thus, for the cavity problem, \autoref{Free} corresponds to the Lifshitz formula, which in the main text is written as Eq.~(4) at zero temperature [$k_{\rm B} T \sum_j\rightarrow\frac{\hbar}{2\pi}\int d\xi$, $\rho(q)=qS/(2\pi)$].

Importantly, \autoref{Free} holds in the absence of spatial dispersion in the dielectric permittivity of the intermirror medium $\varepsilon$. A broader expression can be obtained from the generic form of the light-matter interaction Hamiltonian by explicit integration over the coupling constant in physically justified approximations~\cite{barash1975,barash1989}. We note that the entire derivation of the Lifshitz formula neglects short-wavelength fluctuations (on interatomic scales) and assumes that long-wavelength fluctuations do not induce cavity-size dependence in the medium permittivity
$\varepsilon$; mirror-induced modifications of
$\varepsilon$ are also ignored, consistent with the Lorentz model used.

%Moreover, Barash and Ginzburg provided a general derivation of the free energy of an equilibrium fluctuating EM field in an absorbing medium using the generic form of the light-matter interaction Hamiltonian with the temperature Green's function of the EM field. It turns out that it is possible to explicitly perform integration over the coupling constant in physically justified approximations without specifying the system. For the problem of long-wave fluctuations of the EM field in a cavity, they obtained a more general formula than \autoref{Free}, but which reduces to it under the condition that the permittivity of the inter-mirror medium remains unaffected by the mirrors and exhibits no spatial dispersion (both conditions are within the Lorentz model we used).

\section{III. Overview of Casimir-Lifshitz Calculations}

Here we briefly outline the key features of Casimir-Lifshitz calculations that may be unfamiliar to the general physics community. For a deeper and more comprehensive understanding, see the Refs.~\cite{milonni,mostepanenko,buhmann1,buhmann2,Rodriguez2013,rodriguezRev2011}

%This makes the classical photonics intuition based on the resonance properties and sign of the response functions inapplicable for the Casimir physics. 

\begin{figure}[!tp]
\includegraphics[width=0.3\textwidth]{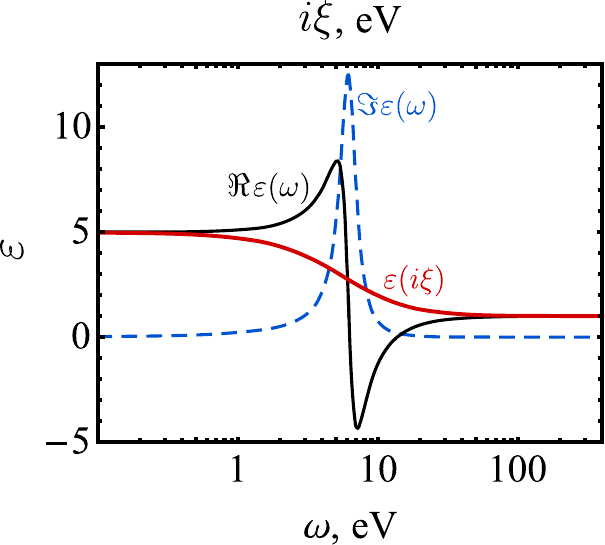}
\caption{\footnotesize{The Lorentzian permittivity function at real and imaginary frequencies, plotted on top of each other (without offset), demonstrating how rotation to the imaginary frequency axis eliminates the material resonance, making the dielectric function monotonic and positive everywhere, while its high and low frequency limits remain unchanged.}}
\label{Fig_eps}
\end{figure}

\textbf{A. Wick rotation}. The interference of various components of the vacuum field in a cavity leads to strong oscillations of the Lifshitz integrand at real frequencies. Although in realistic metal mirrors these oscillations decay at frequencies above the plasma frequency, making calculations at real frequencies possible, subtracting the infinite contribution of free space remains a fundamental condition for obtaining the final result. Additionally, a more general consideration in the Barash-Ginzburg theory presented above implies integration over imaginary frequencies. Similar to other optical linear response functions, the Lifshitz integrand has no singularities in the upper half of the complex frequency plane, and its integral taken over that infinite semicircle vanishes. This enables transforming real-frequency integrals into ones along the positive imaginary axis (Wick rotation). At imaginary frequencies, the Lifshitz integrand becomes monotonic and rapidly decaying, significantly simplifying the calculations. At the same time, the dielectric functions of the constituent materials along the imaginary frequency axis cease to exhibit sharp resonant behavior, and instead become purely real, with values exceeding unity in the entire frequency range (\autoref{Fig_eps}). However, this comes at a price of the necessity to access the response function dispersion in a very broad frequency range. Moreover, due to the factor $\exp\left(-2k_z(q,i\xi) L\right)$ in the Lifshitz formula, the behavior of the response functions at high frequencies becomes important for small cavities $L$, while the behavior at low frequencies always makes a contribution. In particular, the PEC model does not approach unity at high frequencies, as any real material should. As a result, it becomes invalid for sufficiently small $L$ as it leads to a divergence in the van der Waals limit. 

\begin{figure}[!tp]
\includegraphics[width=0.75\textwidth]{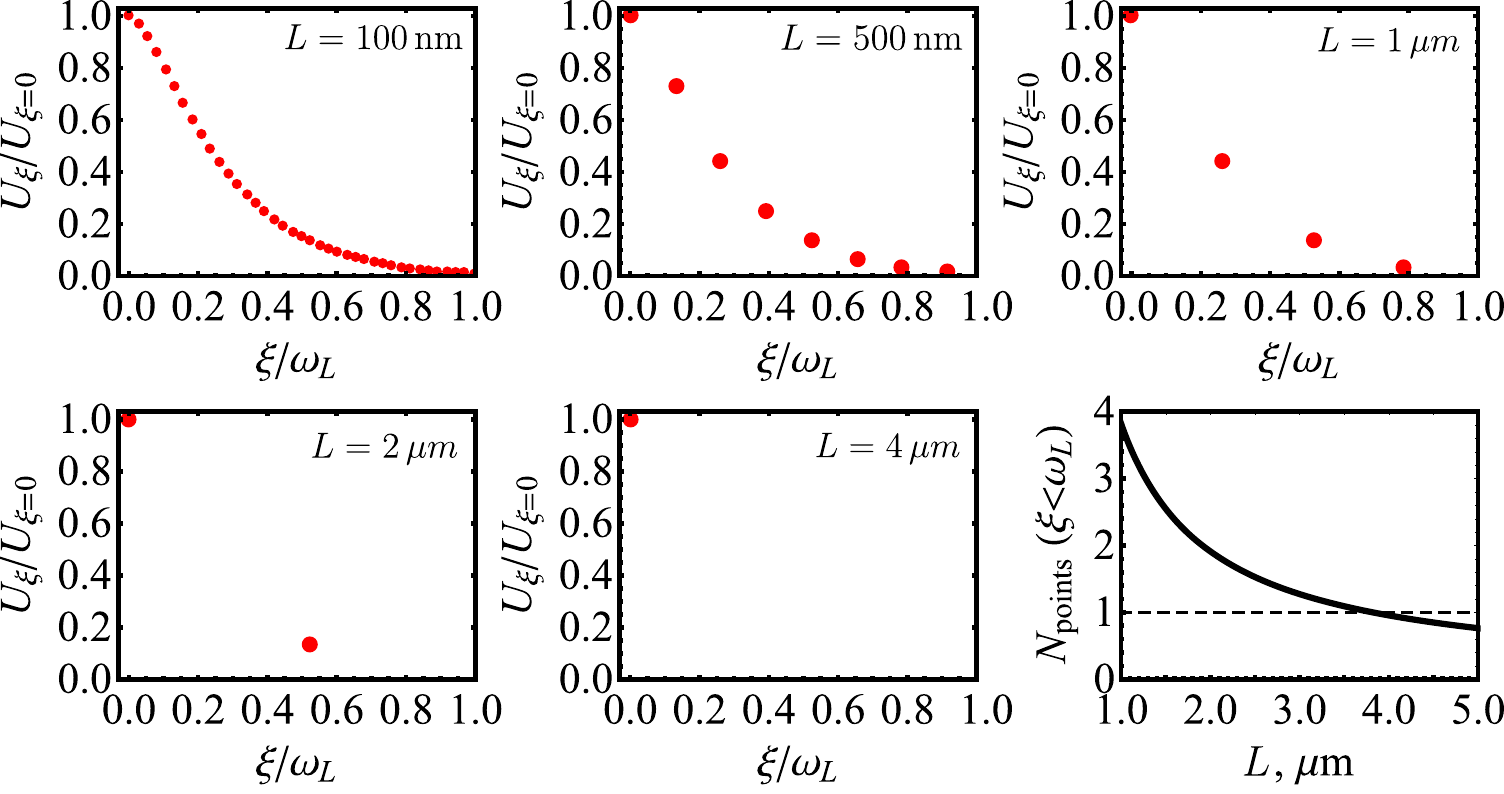}
\caption{\footnotesize{Matsubara frequencies contributing to the normalized Lifshitz integrand with PEC mirrors at room temperature. As the cavity size $L$ increases, their number diminishes, so that at $L>4\mu$m only $\xi=0$ survives, corresponding to the classical regime in which thermal fluctuations completely dominate the quantum ones. The frequency axis is normalized to the main cavity mode $\omega_L=\pi c/L$.}}
\label{Fig_Mats}
\end{figure}

\textbf{B. Finite temperature effect}. At non-zero temperature, the poles of the Bose-Einstein distribution, instead of an integral over the imaginary axis, lead to a sum over Matsubara frequencies $\xi_j=2\pi jk_{\rm B}T/\hbar$ with half the weight at $j = 0$. Again, due to the factor $\exp\left(-2k_z(q,i\xi) L\right)$, the number of Matsubara frequencies contributing to the Lifshitz integrand decreases as $L$ increases (\autoref{Fig_Mats}). This leads to the existence of a crossover from the regime of almost quantum fluctuations at short distances to the classical regime of purely thermal fluctuations at large ones. For metallic mirrors at room temperature, one can neglect the temperature effects at $L\lesssim 1\mu$m; however, already at $L>4\mu$m only the $\xi_j=0$ remains (\autoref{Fig_Mats}), indicating the convergence to a purely thermal (classical) regime. This motivates the use of the zero-temperature Lifshitz formula [Eq.~(4)] for Fig. 2(a),(b) in the main text. Since the dielectric function of the medium between the mirrors enters the Lifshitz formula in the combination $\xi^2\varepsilon(i\xi,\rm g)$, its contribution vanishes entirely at $\xi_j=0$, resulting in the absence of the energy shift in the classical regime. The characteristic distance $L\approx4\mu$m at which this occurs is the same for both PEC and realistic metallic mirrors, since their reflection becomes perfect at $\xi_j=0$.

\textbf{C. Cavity-size dependence}. The single-mode approximation implicitly assumes that the cavity introduces an extra mode absent in the cavity-free limit. As a result, it not only ignores the cavity’s higher-order modes and the full continuum of in-plane wave vectors, but also does not subtract the free-space contribution, which already contains an infinite set of vacuum modes. In reality, the cavity only slightly modifies the vacuum mode structure; therefore, to capture this small cavity-induced change, one must subtract the free-space background. Without this subtraction, the single-mode approximation produces qualitatively incorrect behavior, as illustrated in Fig. 2(b) of the main text. This discrepancy becomes particularly pronounced in the cavity-free limit ($L \rightarrow \infty$), where the single-mode model predicts a constant $\Delta_{\rm pol}/2$, corresponding to the contribution of bulk Hopfield polaritons in free space. 

Another deviation arises from the rapid decay present in the exact Casimir–Lifshitz solution. As is evident from \autoref{wpm}, within the single-mode approximation, the optical cavity and resonant matter parts contribute equally to the polaritonic ZPE. However, the Casimir–Lifshitz energy exhibits a pronounced asymmetry between these contributions: the cavity size enters through a rapidly decaying prefactor ($L^{-3}$ for PEC), whereas the matter term modifies the result only smoothly. This rapidly decaying prefactor, along with the area scaling of the energy, arises from the integration over the in-plane wave vectors and is a hallmark of the Casimir effect, as is evident in the classical Casimir derivation. This causes the exact solution to decay much more rapidly than the single-mode approximation, even if the corresponding free-space subtraction were applied to the latter.

\section{IV. Static screening approximation}
For a PEC cavity, the zero-temperature Lifshitz energy given by Eq.~(4) in the main text can be integrated analytically in the static limit of the permittivity $\varepsilon(i\xi,\mathrm{g})\approx\varepsilon(0,\mathrm{g})$ using first the substitution $p\sqrt\varepsilon \xi/c = \sqrt{q^2+\varepsilon\xi^2/c^2}$ and then the substitution $x = 2pL\sqrt\varepsilon \xi/c$:
\begin{align}
\frac{U_{\rm C}(\mathrm{g},L)}{S} &= 2\frac{\hbar}{4\pi^2}\!\int\limits_{0}^{\infty}\mathit{q}\mathrm{d}q\int\limits_{0}^{\infty} \mathrm{d}\xi
\ln\left(1- e^{-2\sqrt{q^2 + \varepsilon(i \xi, \mathrm{g}) \xi^2 / c^2} L}\right) = \frac{\hbar}{2\pi^2}\!\int\limits_{1}^{\infty} p \mathrm{d}p \int\limits_{0}^{\infty} \mathrm{d}\xi
\frac{\varepsilon(i \xi, \mathrm{g}) \xi^2}{c^2}\ln\left(1- e^{-2 pL \sqrt{\varepsilon(i \xi, \mathrm{g})} \xi/ c}\right)\nonumber
\\
&= \frac{\hbar c}{16\pi^2 L^3}\!\int\limits_{1}^{\infty} \frac{\mathrm{d}p}{p^3} \int\limits_{0}^{\infty} \mathrm{d}x
\frac{x^2\ln{\left(1-e^{-x}\right)}}{\sqrt{\varepsilon(ix,\mathrm{g})}}\approx\frac{\hbar c}{16\pi^2 L^3\sqrt{\varepsilon(0,\mathrm{g})}}\int\limits_{0}^{\infty} \mathrm{d}x\,
x^2\ln\left(1-e^{-x}\right) = -\frac{\hbar c\pi^2}{720 L^3\sqrt{\varepsilon(0,\mathrm{g})}}.\label{SI_SSA}
\end{align} 
Therefore, in the static screening approximation we obtain $U_{\rm C}(\rm g)\approx U_{\rm C}(\rm g=0)/\sqrt{\varepsilon(0,\mathrm{g})}$, which results in Eq.~(5) in the main text. Note that \autoref{SI_SSA} can be viewed as a classical Casimir formula with the speed of light in the medium, resulting in an ``optical'' screening of the Casimir effect.

%\textcolor{red}{Note that \autoref{SI_SSA} can be viewed as a classical Casimir formula with the speed of light in the medium, giving an ``optical'' screening of the Casimir effect, which should be distinguished from electrolyte screening that occurs in some special problems.}

As mentioned before, at imaginary frequencies, the Lorentz permittivity becomes monotonic and positive, but its limits remain the same (\autoref{Fig_eps}): $\varepsilon(0,\mathrm{g})=\varepsilon(i0,\mathrm{g})=1+4\rm g^2/\omega_0^2$ and $\varepsilon(\infty,\mathrm{g})=\varepsilon(i\infty,\mathrm{g})=1$. Although the imaginary zero frequency contains contributions from all real frequencies, $\varepsilon(i0,\mathrm{g})$ ends up being the same as the ordinary static response, which contains contributions from resonances at non-zero frequencies, consistent with the Kramers-Kronig relations.

\section{V. Possible experimental verification}

\begin{figure}[tp]
\includegraphics[width=0.7\textwidth]{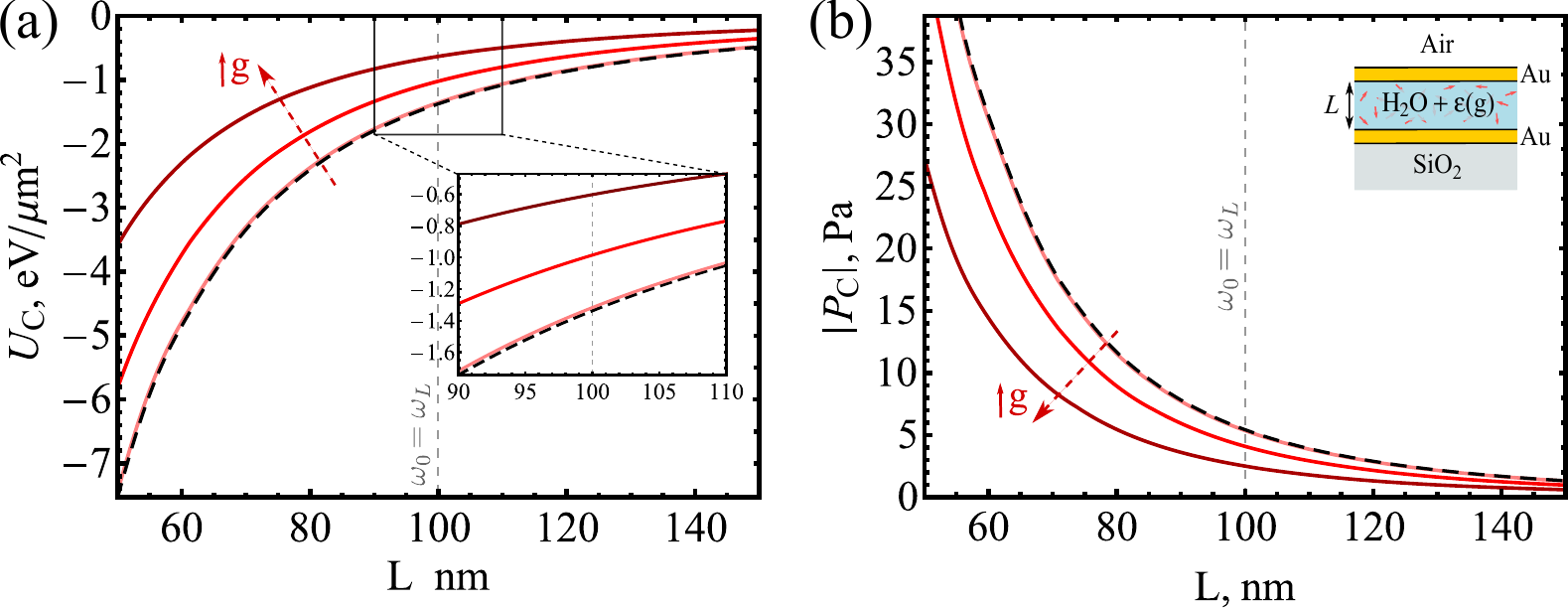}
\caption{\footnotesize{Casimir-Lifshitz energy (a) and pressure (b) for a realistic self-assembled FB cavity formed by 30-nm-gold flakes in an aqueous solution with molecules whose resonant frequency $\omega_0$ matches the fundamental mode of 100-nm-cavity (see the sketch). Different concentrations of molecules lead to $\mathrm{g}/\omega_0$ varying from 0 (black dashed) to 0.1, 0.5, 1 (light to dark red). The inset in (a) shows an enlarged region near the resonance where optical spectra should show polaritonic peaks, but the Casimir energy behaves monotonically, which can be measured by extracting the Casimir energy from the total equilibrium potential of the self-assembled cavities. The absence of pressure ($P=-\partial U/\partial L$) zeros in (b) within the resonance region further confirms the absence of resonances in the energy, which can be probed by the AFM cantilever immersed in liquid.}}
\label{Fig_zoom}

\end{figure}

As shown above, the ground-state energy shift of resonant molecules in an FP cavity can be interpreted as a modification of the corresponding Casimir–Lifshitz energy. This implies that established techniques for measuring the Casimir effect are suitable for testing our theory. Importantly, verifying the basic principle does not require molecules with IR vibrational resonances, as used in typical polaritonic chemistry experiments. Instead, dyes with strong excitonic transitions in the visible or UV range can be used. Practically, this would involve placing molecules which are able to satisfy the strong-coupling condition inside an appropriate FP cavity immersed in an aqueous solution. Standard AFM cantilever measurements~\cite{biggs1994,munday2007,munday2009} of the Casimir pressure in liquids could then be used to qualitatively verify the predicted absence of resonant features. An alternative is the recently proposed approach~\cite{munkhbat2021,kuccukoz2024,hovskova2025} that enables direct determination of the Casimir potential by extracting it from the total equilibrium potential of self-assembled FP cavities. These self-assembled structures -- dimers of flat gold microflakes -- form in colloidal solution through a balance of salt-screened electrostatic repulsion and Casimir-Lifshitz attraction. By tuning the salt concentration, one can produce micrometer-scale dimers of $\sim\,$30-nm-thick flakes separated by $\sim\,$100 -- 200~nm. Such dimers function as good metallic FP cavities~\cite{munkhbat2021}, therefore introducing enough molecules with suitable optical resonances into the liquid should enable the formation of the desired polaritonic cavities. To achieve better control, the bottom gold mirror can be fabricated on a glass substrate, while the top one can then be held in place by a quantum trapping potential inherent to the system~\cite{kuccukoz2024}. In Figure 3 of the main text, we therefore used this system as an example of a realistic polaritonic cavity. In \autoref{Fig_zoom} we reproduce this result, additionally accompanied by the calculated Casimir pressure, emphasizing that any resonance (extremum) in energy would yield zero pressure, since $P=-\partial U/\partial L$. However, our calculations show no sign of such resonance behavior, consistent with a non-resonant effect of polaritons on the ground-state energy. 

For the Lorentz permittivity of the molecules in water, instead of Eq.~(1) from the main text, we used:   
\begin{equation}
 \varepsilon(\omega,\mathrm{g}) =  \varepsilon_{\text{H}_2\text{O}}(\omega) + \frac{4\mathrm{g}^2}{\omega_0^2- \omega^2 - i\omega\gamma},
\end{equation} 
where the water permittivity is taken in the form~\cite{hovskova2025}: $\varepsilon_{\text{H}_2\text{O}}(\omega) = 1 + \frac{\varepsilon_D - 1}{1 - i\omega \tau} +  f\frac{ \omega_{\text{0}}^2}{\omega_{\text{0}}^2 - \omega^2 - i\gamma_0 \omega}$ with  $\varepsilon_D = 77.46$, $\tau =11800$ fs, $ f = 0.843$ , $\omega_{\text{0}} =18.4$ eV, and $\gamma_0$ = 13.5 eV. The gold permittivity was described using the Drude model: $\varepsilon_{\text{Au}}(\omega) = 1 - \frac{\omega_p^2}{\omega(\omega + i\gamma_D)}$
with $\omega_p$ = 8.9 eV and $\gamma_D$ = 0.0357 eV. For a realistic structure, we obtain the relative change of $\Delta U/U(0)\approx\Delta P/P(0)\approx1\%$ at the USC parameters ($\mathrm{g}=0.1\omega_0$) instead of $2\%$ expected from the static screening approximation without water. Although the predicted effect is small, these values are not too distant from the current limits of Casimir pressure measurement accuracy; therefore, it is reasonable to expect that they will become accessible in the foreseeable future, for example, through an improved self-assembly approach~\cite{hovskova2025}.

\end{document}